\documentclass[pdftex,twocolumn,epjc3]{svjour3}          

\RequirePackage[T1]{fontenc}

\smartqed  

\RequirePackage{graphicx}
\RequirePackage{amsmath}
\RequirePackage{flushend}
\RequirePackage[numbers,sort&compress]{natbib}
\RequirePackage[colorlinks,citecolor=blue,urlcolor=blue,linkcolor=blue]{hyperref}

\journalname{Eur. Phys. J. C}

\begin{document}

\title{The Tsallis Distribution at Large Transverse Momenta}

\author{M. D. Azmi\thanksref{e1,addr1} \and J. Cleymans\thanksref{e2,addr2} }

\thankstext{e1}{e-mail: danish.hep@gmail.com}   
\thankstext{e2}{e-mail: jean.cleymans@uct.ac.za}

\institute{HEP Lab, Physics Department, Aligarh Muslim University, Aligarh - 202002, India\label{addr1}
          \and
          UCT - CERN Research Centre, Physics Department, University of Cape Town, Rondebosch - 7701, South Africa \label{addr2} }

\date{Received: date / Accepted: date}

\maketitle

\begin{abstract}
The charged particle transverse momentum ($p_T$) spectra measured by the ATLAS and CMS collaborations for proton - proton collisions 
at $\sqrt{s}$ = 0.9 and 7 TeV have been studied using Tsallis thermodynamics. A thermodynamically consistent form of the 
Tsallis distribution is used for fitting the transverse momentum spectra at mid-rapidity. It is found that the fits based on 
the proposed distribution are working over 14 orders of magnitude with $p_T$ values up to 200 GeV/c
and gives a value of $\chi^2/NDF$ of 0.52/24 for the  CMS data at 7 TeV. 
The values for  $\frac{dN}{dy}|_{y=0}$ as a function of centre-of-mass energy obtained using Tsallis distribution 
for both ATLAS and CMS data sets are presented and discussed.     
\end{abstract}

\section{Introduction}
It is  well known that the Tsallis distribution gives excellent fits to the transverse momentum distributions observed at
the Relativistic Heavy Ion Collider (RHIC)~\cite{STAR,PHENIX1,PHENIX2} and at the Large Hadron 
Collider (LHC)~\cite{ALICE,CMS1,CMS2,ATLAS,ALICE2} with only three parameters, $q$, $T$ and $dN/dy$ (or, alternatively, a 
volume $V$~\cite{ryb, cleymans1, azmi-cleymans}). The parameter $q$ is referred as Tsallis parameter and 
discussed elsewhere~\cite{wilk4} in detail. The parameter $T$ obeys the standard thermodynamic relation, 
\begin{equation*}
\left. T = \frac{\partial E}{\partial S}\right |_{N, V},  
\end{equation*}
however, the entropy used in the  above equation is the  Tsallis entropy~\cite{tsallis1} not the standard Boltzmann-Gibbs (BG) entropy. The parameter $V$ is not necessarily related to a volume deduced from other models like HBT calculations.

It was recently shown that these fits extend to values~\cite{CMS3} of $p_T$ up to 200 GeV/c~\cite{wilk2,tsallis-poland} . This is unexpected because in this kinematic range hard scattering processes become important~\cite{wilk3}. A description of the high $p_T$ results has been discussed in \cite{Urmossy} where a model using a combination of Tsallis at low $p_T$ and QCD hard scattering at high $p_T$ was considered. The present analysis shows that the Tsallis distribution describes measurements up to the highest $p_T$ using the same Tsallis parameters as those obtained at low $p_T$.

A power law based on the Tsallis distribution~\cite{tsallis1} is used to fit the $p_T$ spectra of charged particles measured by the ATLAS and CMS collaborations.  The ATLAS collaboration has reported the transverse momentum in an inclusive phase space region taking into account at least two charged particles in the kinematic range $|\eta|$ < 2.5 and $p_T$ > 100 MeV~\cite{ATLAS}. 
The CMS collaboration has presented the differential transverse momentum distribution covering a $p_T$ range up to 200 GeV/c, the 
largest range ever measured in a colliding beam experiment~\cite{CMS3}. 
The results for the charged particle multiplicities are consistent with those of ATLAS measurements. 

Moreover, the results are compared to those obtained in~\cite{wilk2,tsallis-poland,wilk3,wilk5,wilk6} where very good fits to transverse momentum distributions were presented. We confirm the quality of the fits but obtain  different values of the  parameters albeit using a different version of the Tsallis model.

\section{Tsallis Distribution}
Power law distributions have been widely applied in high energy physics~\cite{STAR,PHENIX1,PHENIX2,ALICE,CMS1,CMS2,ATLAS,ALICE2} to describe the transverse momentum spectra of secondary particles produced in proton - proton collisions. With the advent of the LHC, transverse 
momenta of hundreds of GeV have been measured. It is of great interest to choose a distribution which describes such high $p_T$ measurements very well. A well suited form of the Tsallis distribution is discussed in this paper and our main criterium for choosing such a distribution was thermodynamic consistency which has not always been implemented fully~\cite{worku1,worku2,conroy}. 

It is well established that there are numerous physical systems under which BG statistics encounters many difficulties. In particular, when analysing the $p_T$ spectra of hadrons it is found that spectra decrease far slower than predicted by BG statistics, and appear to follow some power law at high $p_T$. The Tsallis distribution was first proposed 
about twenty - seven years~\cite{tsallis1} ago as a generalization of the BG distribution. 

The Tsallis form of the BG distribution is given by:
\begin{align}
    f(E)  \equiv \exp_{q}\left(-\frac{E-\mu}{T}\right) \label{eq1}, 
\end{align}
where $E$ is the energy and $\mu$ is the chemical potential and 
the function $\exp_{q}(x)$ is usually defined as:
\begin{equation}
\exp_{q}(x) \equiv
  \begin{cases}
    [1+(q-1)x]^{1/(q-1)}       & \quad \text{if }  x > 0\\
    [1+(1-q)x]^{1/(1-q)}       & \quad \text{if }  x \le 0\\
  \end{cases}
\end{equation}
in the  limit where $q \rightarrow$ 1 it reduces to the standard exponential:
\begin{eqnarray*}
\lim_{q \to 1} \exp_q(x) \rightarrow \exp(x).
\end{eqnarray*}

The expressions of the relevant thermodynamic quantities like entropy, $S$, energy density, $\epsilon$ (= E/V), pressure, $P$, and 
particle number, $N$, are given below using the $q$-logarithm function, 
\begin{equation*}
{\rm ln}_q(x) \equiv \frac{x^{1-q} - 1}{1 - q},
\end{equation*}
as integrals over  the function defined in eq. (\ref{eq1})~\cite{worku1,worku2,azmi-cleymans}:
\begin{gather}
S = - gV\int\frac{d^3p}{(2\pi)^3}\left[f^{q}{\rm ln}_{q}f - f\right],\\
\epsilon = g\int\frac{d^3p}{(2\pi)^3}~E~f^{q},\\
P = g\int\frac{d^3p}{(2\pi)^3}\frac{p^{2}}{3E}~f^{q},\\
N = gV\int\frac{d^3p}{(2\pi)^3} f^{q} \label{eq7},
\end{gather}
where $g$ is the degeneracy factor and $V$ is the volume. 
The Tsallis distribution used in this paper is thermodynamically consistent which means that the following relations, derived from the differential form of the first and second laws of thermodynamics, are satisfied~\cite{worku2}:
\begin{align}
T  &= \left.\frac{\partial \epsilon}{\partial s}\right|_n,   &   \mu  &= \left.\frac{\partial \epsilon}{\partial n}\right|_s, \nonumber\\
n  &= \left.\frac{\partial P}{\partial \mu}\right|_T,       &   s  &= \left.\frac{\partial P}{\partial T}\right|_\mu,
\label{consistency}
\end{align}
where $s$ and $n$ are the entropy and particle number densities respectively. 

The corresponding momentum distribution deduced from eq. (\ref{eq7}) is given by:
\begin{equation}
E \frac{d^{3}N}{dp^3} = gVE \frac{1}{(2\pi)^3}\left[1 + (q-1)\frac{E-\mu}{T}\right]^{-\frac{q}{q-1}},
\end{equation}
which in terms of the rapidity, $y$, and transverse mass, $m_T$, variables ($E = m_{T}\cosh y$) at mid rapidity and 
for $\mu$  = 0 becomes~\cite{azmi-cleymans,worku1}:  
\begin{equation}
\left.\frac{d^{2}N}{dp_{T}dy}\right|_{y = 0} = gV\frac{p_{T}m_{T}}{(2\pi)^2}\left[1 + (q - 1) \frac{m_T}{T}\right]^{-\frac{q}{q-1}} \label{eq9}.
\end{equation}

The relationship between the above  parameterization  and the one used by the RHIC and LHC 
collaborations~\cite{STAR,PHENIX1,PHENIX2,ALICE,CMS1,CMS2,ATLAS,ALICE2} has been discussed in~\cite{cleymans1}.

Additionally,   comparison with high-energy data has been done in~\cite{wilk2,tsallis-poland,wilk3,wilk5,wilk6}, using
a simplified form:
\begin{equation}
\left.\frac{d^{2}N_{ch}}{dp_Tdy}\right|_{y = 0} = A\left[1 + (q - 1)\frac{p_T}{T}\right]^{-\frac{1}{q - 1}}
\end{equation} 
where A is a normalization factor and $T$ is a parameter used to fit the transverse momentum distribution and is not 
related to a temperature in the thermodynamic sense as in eq. (\ref{eq9}). 

Furthermore, integration of eq. (\ref{eq9}) over the transverse momentum leads to~\cite{ryb}:
\begin{gather*}
\left.\frac{dN}{dy}\right|_{y = 0} = \frac{gV}{(2\pi)^2} \int_0^\infty p_{T}dp_{T}m_{T} \left[1 + (q - 1) \frac{m_T}{T}\right]^{-\frac{q}{q-1}}
\end{gather*}
\begin{gather}
= \frac{gVT}{(2\pi)^2} \left[\frac{(2 - q)m^{2} + 2mT + 2T^{2}}{(2 - q)(3 - 2q)}\right]\left[1 + (q - 1) \frac{m}{T}\right]^{-\frac{1}{q-1}}, \label{eq11}
\end{gather}
where $m$ is the rest mass. 

Using the values of parameters $q$, $T$ and $V$ obtained from Tsallis fit, using eq. (\ref{eq9}), above equation enable us to obtain the values of $dN/dy$ for pions, kaons and protons. Their sum give us the values of $dN/dy$ for charged particles.  

\section{Fit Details}
The charged particle transverse momentum spectra measured by the ATLAS and CMS detectors in proton - proton 
collisions at LHC energies are fitted using eq. (\ref{eq9}). 

It is well established that the charged particle $p_T$ spectra is composed of contribution from pions, kaons and protons. Hence, a sum of three Tsallis distributions for pions, $\pi^{+}$'s, kaons, $K^{+}$'s and protons, $p$, has been applied to fit the charged particle $p_T$ spectra in following way:
\begin{align}
\left.\frac{d^2N_{ch}}{dp_Tdy}\right|_{y = 0} &= 2p_T\frac{V}{(2\pi)^{2}}\sum_{i = 1}^3g_{i}m_{T,i}\left[1 + (q - 1) \frac{m_{T,i}}{T}\right]^{-\frac{q}{q-1}} \label{eq12}
\end{align}
where $i = \pi^{+}, K^{+}, p$.  The additional factor of 2 on the right-hand side of eq. (\ref{eq12}) takes into account the negatively
charged particles $\pi^{-}, K^{-}, \bar{p}$. The relative weights between particles were determined by the corresponding degeneracy factors and given by $g_{\pi^{+}}$ = $g_{K^{+}}$ = 1 and $g_p$ = 2. The square root of the sum of the squares of the statistical and systematic errors (quadrature) was taken into account for fitting the data. 

\begin{figure} 
\begin{minipage}{\columnwidth}
\centering
\raisebox{10pt}[85mm][0mm]{\includegraphics[width=1.05\textwidth]{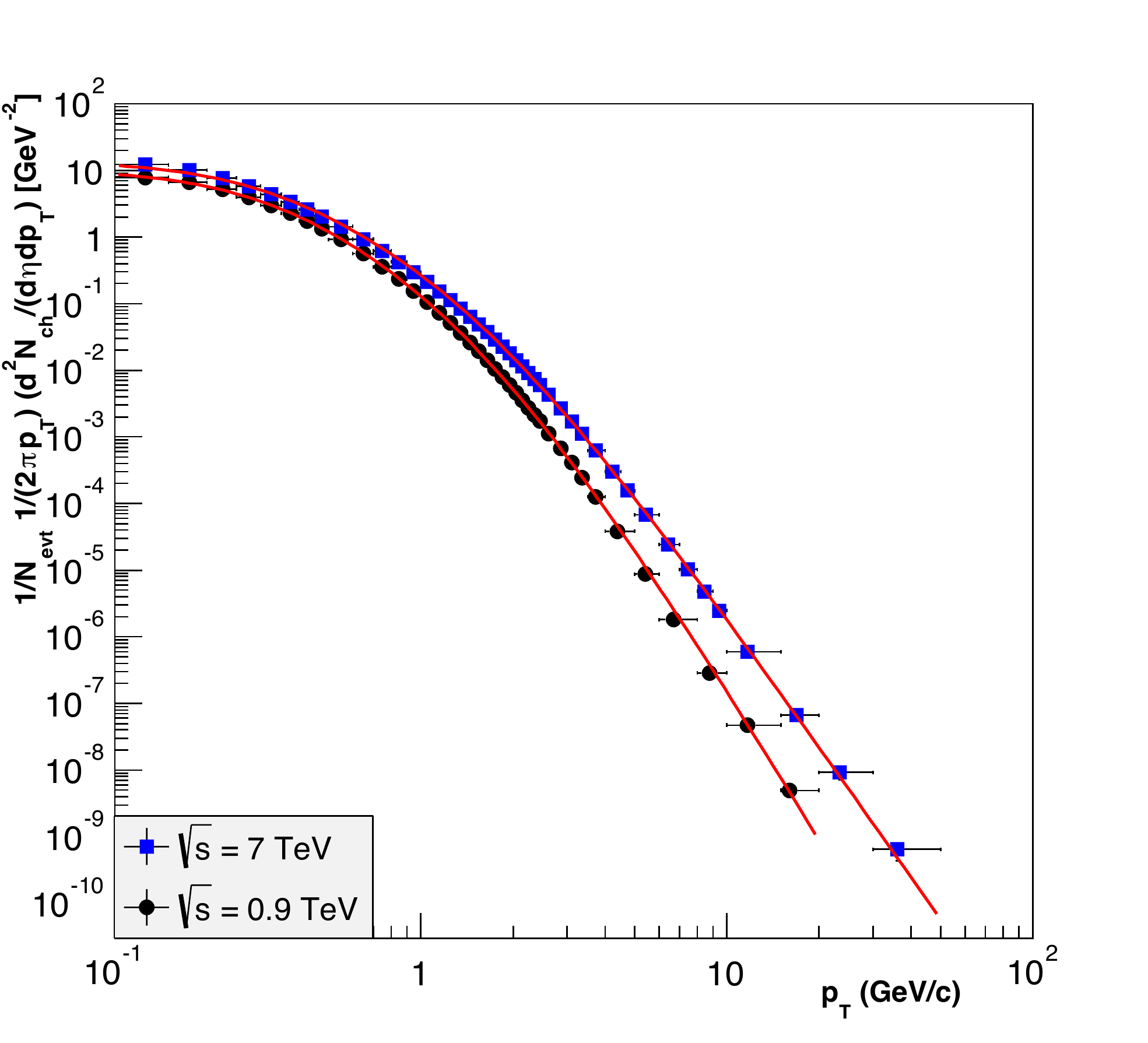}}
\end{minipage}
\caption{Charged particle multiplicities as a function of the transverse momentum measured by the ATLAS detector for events with $n_{ch} \ge 2$, $p_T$ > 100 MeV and $|\eta|$ < 2.5 at $\sqrt{s}$ = 0.9 and 7 TeV in proton - proton collisions~\cite{ATLAS} fitted with Tsallis distribution.}
\label{atlas1}
\end{figure}

\begin{figure}
\begin{minipage}{\columnwidth}
\centering
\raisebox{10pt}[85mm][0mm]{\includegraphics[width=0.95\textwidth]{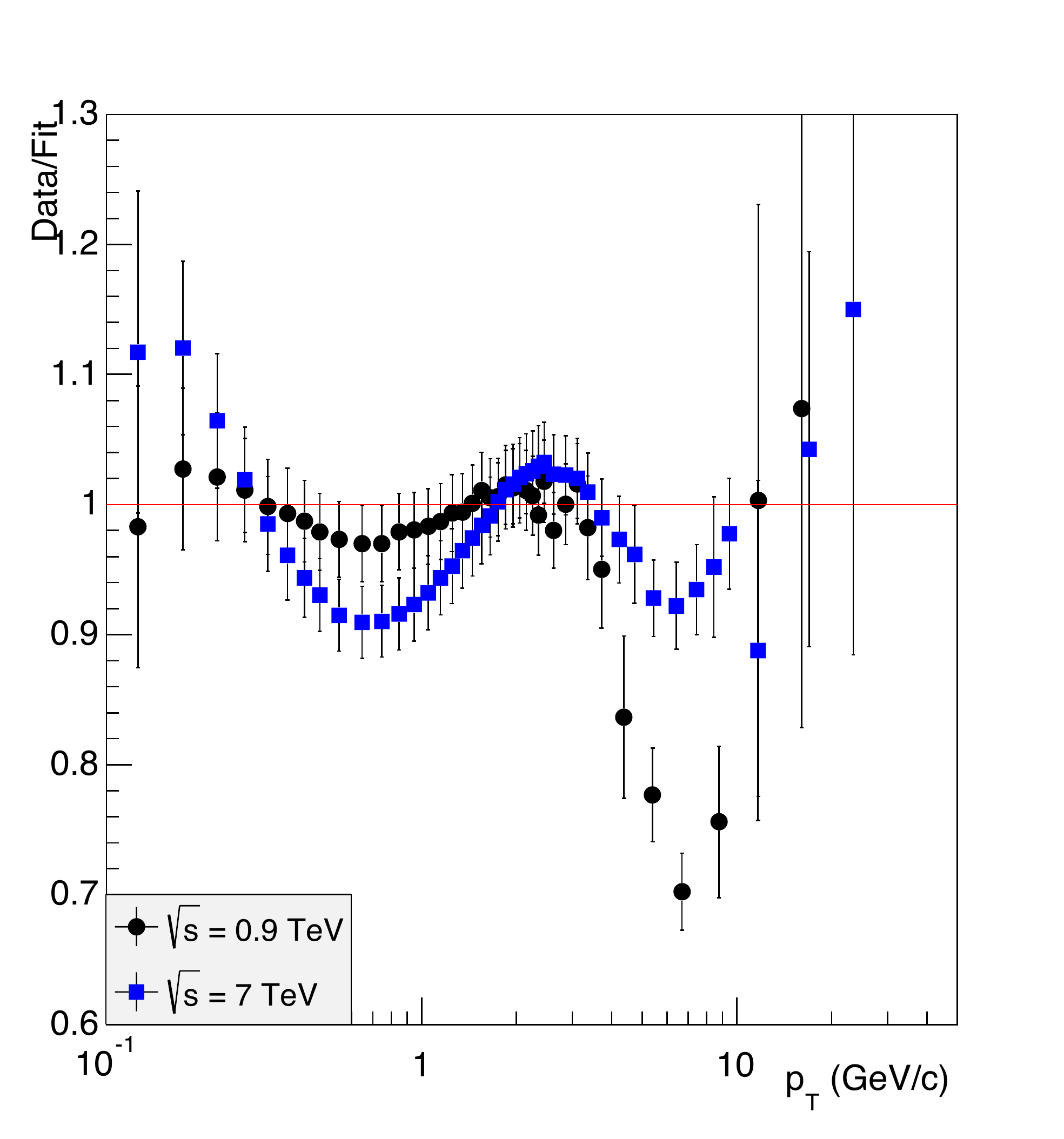}}
\end{minipage} 
\caption{Ratios of the data to the fit values measured for ATLAS detector at $\sqrt{s}$ = 0.9 and 7 TeV.}
\label{atlas2}
\end{figure}

The charged particle multiplicity as a function of the transverse momentum measured by the ATLAS collaboration for events 
with $n_{ch} \ge 2$, $p_T$ > 100 MeV and $|\eta|$ < 2.5 at $\sqrt{s}$ = 0.9 and 7 TeV in proton - proton collisions are shown 
in figure~\ref{atlas1}. The parameterisation given in eq. (\ref{eq12}) has been used to fit the spectra. 
In order to show the quality of the fits in more detail, Figure~\ref{atlas2} shows the ratios of the data to the 
fit values at $\sqrt{s}$ = 0.9 and 7 TeV for the ATLAS data. The largest deviations occur at the lower energy 0.9 TeV while the
7 TeV data show  deviations from the Tsallis fits which are at most in the 10 \% range.  

\begin{figure}
\begin{minipage}{\columnwidth}
\centering
\raisebox{10pt}[88mm][0mm]{\includegraphics[width=1.02\textwidth]{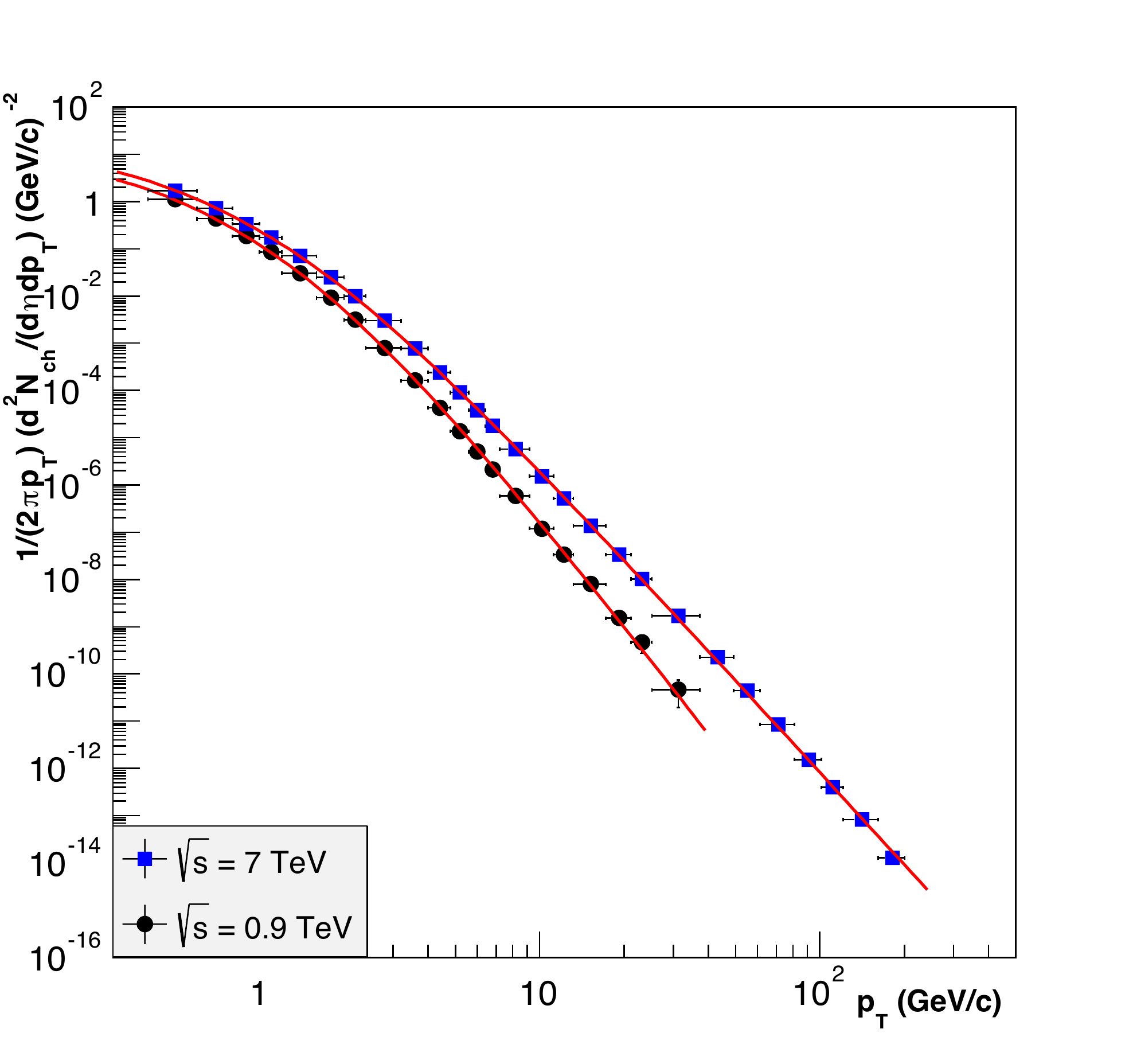}}
\end{minipage} 
\caption{Charged particle differential transverse momentum yields measured within $|\eta|$ < 2.4 by the CMS detector in proton - proton collisions at $\sqrt{s}$ = 0.9 and 7 TeV~\cite{CMS3} fitted with Tsallis distribution.}
\label{cms1}
\end{figure}

\begin{figure}
\begin{minipage}{\columnwidth}
\centering
\raisebox{10pt}[92mm][0mm]{\includegraphics[width=0.95\textwidth]{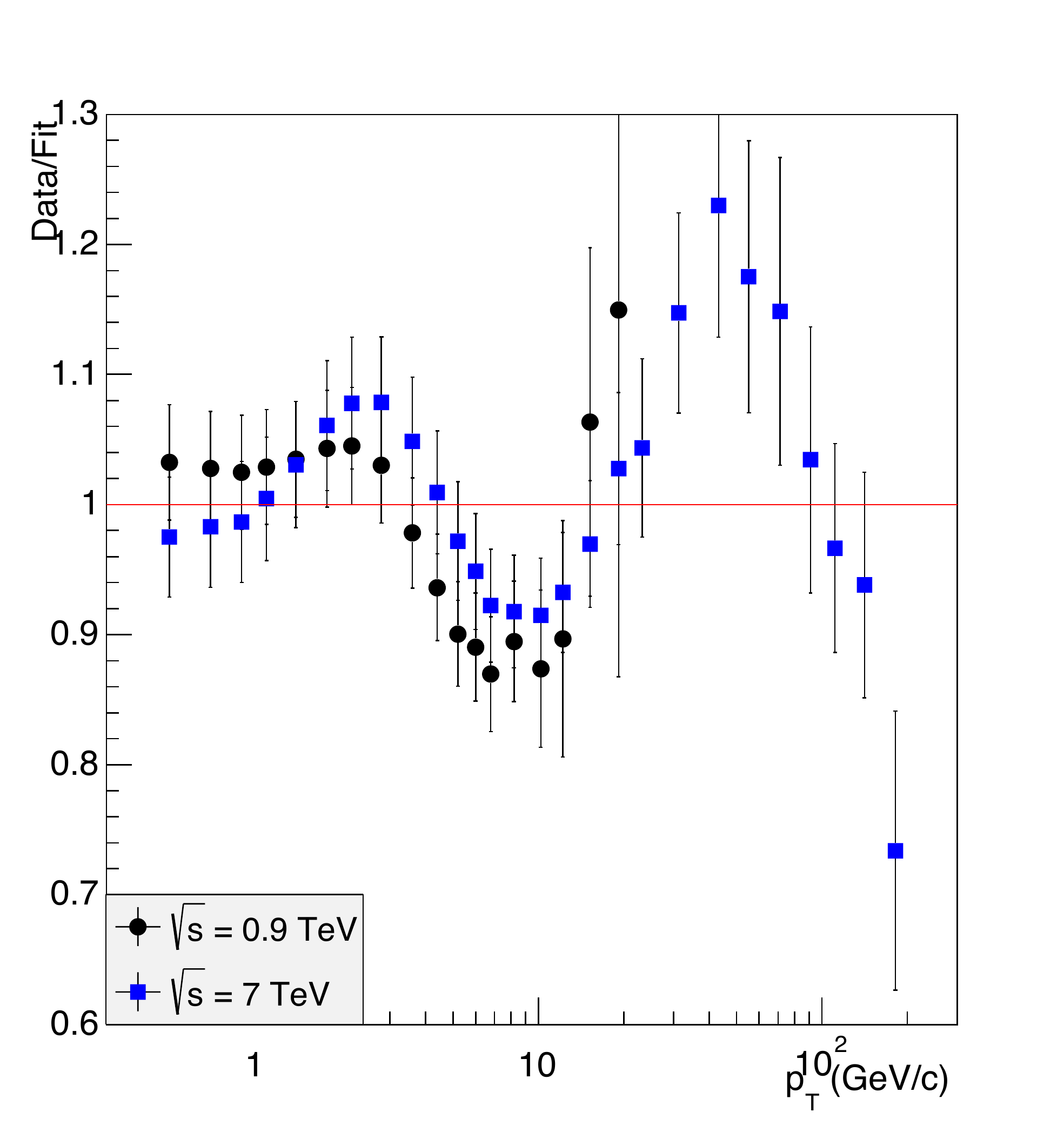}}
\end{minipage} 
\caption{Ratios of the data to the fit values measured for CMS detector at $\sqrt{s}$ = 0.9 and 7 TeV.}
\label{cms2}
\end{figure}

The charged particle differential transverse momentum yields in proton - proton collisions at $\sqrt{s}$ = 0.9 and 7 TeV have been 
measured within $|\eta|$ < 2.4 by the CMS detector. The CMS measurements are described by the Tsallis fit and displayed in figure~\ref{cms1}. 
The ratios of the data to the Tsallis fit at $\sqrt{s}$ = 0.9 and 7 TeV are shown in figure~\ref{cms2}. 
Again, in order to sow the quality of the fits in more detail, Figure~\ref{cms2} shows the ratios of the data to the 
fit values at $\sqrt{s}$ = 0.9 and 7 TeV for the CMS data. The largest deviations occur at the highest values of $p_T$
where also the error bars are largest.  

\section{Results and Discussions}
A thermodynamically consistent form of the Tsallis distribution, given in eq. (\ref{eq9}), has been used to fit the transverse 
momentum spectra of the charged particle measured by both ATLAS and CMS collaborations in proton - proton collisions 
at $\sqrt{s}$ = 0.9 and 7 TeV. It is observed from figure~\ref{atlas1} and figure~\ref{cms1} that the Tsallis distribution fits well the measured $p_T$ spectra, albeit 
that the kinematical conditions of the data of both detectors were not identical. 
It is clear that the proposed form of the Tsallis distribution fits not only the low $p_T$ but covers a wide range of $p_T$ spectra upto 200 GeV/c.  

The ratios of the experimental data over the fit values, depicted in figure~\ref{atlas2} and figure~\ref{cms2}, shows an 
intriguing log-periodic oscillation as a function of the transverse momentum not only for high $p_t$ data sets~\cite{CMS3} 
but also for low $p_T$ data sets~\cite{ATLAS}. This was first noticed in~\cite{wilk5,wilk6} for CMS data sets only. 
The origin of such oscillations is a matter of investigation in order to draw some useful physics out of it. 

The values of the fit parameters, $q$, $T$ and the corresponding $\chi^2/NDF$ obtained from the fits of the $p_T$ spectra 
are given in table~\ref{par1}. 
For the highest values of $p_T$, reaching 200 GeV/c,  a value of $\chi^2/NDF\approx 0.52/24$ for the  CMS data is obtained. 

Instead of parameter $V$, a parameter $R \equiv [V \frac{3}{4\pi}]^{1/3}$, referred as 
radius, is listed in the table. It has been found that the values of $q$ increases slowly but clearly with 
increase in the center-of-mass energy. 
The thermodynamically consistent version leads to a temperature $T \approx 75$ MeV  at both  values of $\sqrt{s}$. 
It can be observed that the parameter $R$ $\approx$ 4.5 fm and 5.0 fm for $\sqrt{s}$ = 0.9 and 7 TeV respectively for both ATLAS and CMS data sets. These results are in agreement with the previous analysis done for lower $p_T$ data~\cite{azmi-cleymans,cleymans1}.   

Table~\ref{par2} shows the values of $\left.\frac{dN}{dy}\right|_{y=0}$ for pions, kaons and protons and total 
charged particles evaluated using eq. (\ref{eq11}). 
The $\left.\frac{dN_{ch}}{d\eta}\right|_{\eta=0}$ values measured by the ATLAS detector is also present. It can be found that the production of pions are less and more kaons and protons are produced at higher energies. The charged particle multiplicity found for ATLAS using Tsallis fit are close to the measured values. An estimate of the charged particle multiplicity is presented for CMS which can be verified by the CMS collaboration.
  
\section{Conclusions}
We find it quite remarkable that the  transverse momentum distributions measured up to 200 GeV/c in $p_T$ can be described 
consistently over 14 orders of magnitude by a straightforward Tsallis distribution as presented in eq. (\ref{eq12}). 
The values of $\chi^2/NDF$ are listed in table 1 and for the highest values of beam energy and $p_T$ it is found to be 0.52/24 for the CMS data. 
The distribution used is part of a consistent thermodynamic description satisfying the basic consistency conditions given in eq. (\ref{consistency}). 

In conclusion, we have shown that the proposed version of the Tsallis distribution only fits the transverse momentum spectra with 
minimal values  of $\chi^2$ values and an estimate of the particle multiplicities and particle ratio has been obtained.

\begin{table*}
\caption{Values of the $q$, $T$ and $R$ parameters and $\chi^2/NDF$ obtained using eq. (\ref{eq9}) to fit the $p_T$ spectra measured by the ATLAS~~\cite{ATLAS} and CMS~\cite{CMS3} detectors.}
\label{par1}
\begin{tabular*}{\textwidth}{@{\extracolsep{\fill}}llllll@{}}
\hline
{\bf Experiment} & \multicolumn{1}{c}{\bf $\sqrt{s}$ (TeV)} & \multicolumn{1}{c}{\bf $q$} & \multicolumn{1}{c}{\bf $T$ (MeV)} & \multicolumn{1}{c}{\bf $R$ (fm)} & \multicolumn{1}{c}{\bf $\chi^2/NDF$} \\
\\
\hline
ATLAS & 0.9 & 1.129 $\pm$ 0.005 & 74.21 $\pm$ 3.55 & 4.62 $\pm$ 0.29 & 0.657503/36 \\
ATLAS & 7 & 1.150 $\pm$ 0.002 & 75.00 $\pm$ 3.21 & 5.05 $\pm$ 0.07 & 4.35145/41 \\
\hline
CMS & 0.9 & 1.129 $\pm$ 0.003 & 76.00 $\pm$ 0.17 & 4.32 $\pm$ 0.29 & 0.648806/17 \\
CMS & 7 & 1.153 $\pm$ 0.002 & 73.00 $\pm$ 1.42 & 5.04 $\pm$ 0.27 & 0.521746/24 \\
\hline
\end{tabular*}
\end{table*}

\begin{table*}
\caption{Values of $\left.\frac{dN}{dy}\right|_{y=0}$ for pions, kaons, protons and total charged particles obtained 
using eq. (\ref{eq11}) and $\left.\frac{dN_{ch}}{d\eta}\right|_{\eta=0}$ measured by ATLAS detector.}
\label{par2}
\begin{tabular*}{\textwidth}{@{\extracolsep{\fill}}lllllll@{}}
\hline
{\bf Experiment} & \multicolumn{1}{c}{\bf $\sqrt{s}$ (TeV)} & \multicolumn{1}{c}{\bf{$\left.\frac{dN^{\pi^{+}+\pi^{-}}}{dy}\right|_{y=0}$}} & \multicolumn{1}{c}{\bf$\left.\frac{dN^{K^{+}+K^{-}}}{dy}\right|_{y=0}$} & \multicolumn{1}{c}{\bf$\left.\frac{dN^{p+\bar{p}}}{dy}\right|_{y=0}$} & \multicolumn{1}{c}{\bf{$\left.\frac{dN}{dy}\right|_{y=0}$}} & \multicolumn{1}{c}{\bf$\left.\frac{dN_{ch}}{d\eta}\right|_{\eta=0}$}\\
\\
\hline
ATLAS & 0.9 & 2.84 & 0.76 & 0.32 & 3.92 & 3.85 $\pm$ 0.18 \\
ATLAS & 7 & 4.27 & 1.37 & 0.72 & 6.36 & 6.25 $\pm$ 0.30 \\
\hline
CMS & 0.9 & 2.52 & 0.70 & 0.30 & 3.52 & \\
CMS & 7 & 3.92 & 1.24 & 0.66 & 5.82 & \\
\hline
\end{tabular*}
\end{table*}

\end{document}